\documentclass[journal=apchd5,manuscript=article]{achemso}

\usepackage{chemformula} 
\usepackage[T1]{fontenc} 
\usepackage{graphicx}%
\usepackage[margin=1cm,font=small,labelfont=bf,format=plain,labelsep=period, justification=justified,figureposition=below, tableposition=top, skip=5pt]{caption}
\usepackage{multirow}%
\usepackage{amsmath,amssymb,amsfonts}%
\usepackage{mathtools}
\usepackage{amsthm}%
\usepackage{mathrsfs}%
\usepackage[title]{appendix}%
\usepackage{xcolor}%
\usepackage{textcomp}%
\usepackage{manyfoot}%
\usepackage{booktabs}%
\usepackage{algorithm}%
\usepackage{algorithmicx}%
\usepackage{algpseudocode}%
\usepackage{listings}%
\usepackage{geometry}
\author{Qingzhong Deng}
\affiliation{imec, Kapeldreef 75, 3001 Leuven, Belgium}
\email{qingzhong.deng@imec.be}
\author{Alaa Elshazly}
\affiliation{imec, Kapeldreef 75, 3001 Leuven, Belgium}
\alsoaffiliation{Photonics Research Group, Department of Information Technology, Ghent University-imec, 9000 Ghent, Belgium}
\author{Mehmet Oktay}
\affiliation{imec, Kapeldreef 75, 3001 Leuven, Belgium}
\alsoaffiliation{Photonics Research Group, Department of Information Technology, Ghent University-imec, 9000 Ghent, Belgium}
\author{Jeroen De Coster}
\affiliation{imec, Kapeldreef 75, 3001 Leuven, Belgium}
\author{Rafal Magdziak}
\affiliation{imec, Kapeldreef 75, 3001 Leuven, Belgium}
\author{Chiara Marchese}
\affiliation{imec, Kapeldreef 75, 3001 Leuven, Belgium}
\author{Guy Lepage}
\affiliation{imec, Kapeldreef 75, 3001 Leuven, Belgium}
\author{Dieter Bode}
\affiliation{imec, Kapeldreef 75, 3001 Leuven, Belgium}
\author{Hakim Kobbi}
\affiliation{imec, Kapeldreef 75, 3001 Leuven, Belgium}
\author{Hemant Kumar Tyagi}
\affiliation{imec, Kapeldreef 75, 3001 Leuven, Belgium}
\author{Javad Rahimi Vaskasi}
\affiliation{imec, Kapeldreef 75, 3001 Leuven, Belgium}
\author{Marko Ersek Filipcic}
\affiliation{imec, Kapeldreef 75, 3001 Leuven, Belgium}
\author{Huseyin Sar}
\affiliation{imec, Kapeldreef 75, 3001 Leuven, Belgium}
\author{Maumita Chakrabarti}
\affiliation{imec, Kapeldreef 75, 3001 Leuven, Belgium}
\author{Dimitrios Velenis}
\affiliation{imec, Kapeldreef 75, 3001 Leuven, Belgium}
\author{Peter Verheyen}
\affiliation{imec, Kapeldreef 75, 3001 Leuven, Belgium}
\author{Philippe Absil}
\affiliation{imec, Kapeldreef 75, 3001 Leuven, Belgium}
\author{Filippo Ferraro}
\affiliation{imec, Kapeldreef 75, 3001 Leuven, Belgium}
\author{Yoojin Ban}
\affiliation{imec, Kapeldreef 75, 3001 Leuven, Belgium}
\author{Joris Van Campenhout}
\affiliation{imec, Kapeldreef 75, 3001 Leuven, Belgium}

\title{Silicon Ring Based 64$\times$100 GHz Wavelength Division Multiplexing filter}

\begin{document}

\begin{abstract}
Silicon-based wavelength division multiplexing (WDM) filters are essential for scaling optical communication capacity in data centers and telecommunications networks.
However, extending silicon WDM systems beyond 32 channels with 100 GHz spacing poses significant challenges due to limitations in conventional filter architectures.
Here we present the first silicon 64$\times$100 GHz WDM filter by introducing a novel ring-Mach-Zehnder interferometer (MZI) cascade architecture.
Our design utilizes third-order polynomial interconnected circular (TOPIC) bends to construct low-loss half-ring waveguides, facilitating an MZI configuration where the arm length difference is determined entirely by half of the ring structure.
This approach ensures precise alignment between the MZI interference peaks and the ring resonator wavelengths, with the MZI FSR being exactly double that of the ring, eliminating the need for dynamic tuning between the MZI and the ring.
We demonstrate the concept through a 16$\times$400 GHz WDM filter with insertion loss of 1.3$\pm$0.6 dB and channel isolation $\geq$14.3 dB.
The 64$\times$100 GHz implementation, realized using a 4-channel interleaver followed by four 16$\times$400 GHz WDM filter, achieves insertion loss of 3.2$\pm$1.1 dB and channel isolation $\geq$10.7 dB.
This work opens new possibilities for high-density silicon photonic WDM systems, addressing the growing bandwidth demands of artificial intelligence and machine learning applications.

\end{abstract}

\section{Introduction}

Wavelength division multiplexing (WDM) serves as the enabling technology for scaling optical communication capacity in silicon photonics, particularly as data centers and telecommunications networks demand ever-increasing throughput driven by artificial intelligence and machine learning workloads.
Increasing the number of WDM channels over a larger spectral range enables higher communication capacity.
However, silicon WDM systems face significant challenges in WDM filter design to extend beyond a spectral range of 3.2 THz with high channel counts.
Silicon WDM filters are typically implemented using cascaded Mach-Zehnder interferometers (MZIs), arrayed waveguide gratings (AWGs), echelle gratings, or ring resonators.
The cascaded MZI approach is demonstrated up to 32$\times$50 GHz WDM, which has suffered from a high insertion loss of 9.8 dB~\cite{CrosstalkFree32_JOLT2023a}.
The AWG and echelle grating designs have struggled to go beyond 32$\times$100 GHz WDM, including 40$\times$100 GHz~\cite{LowLossHigh_2012a, CompactSingleChip_OE2011a} and 45$\times$100 GHz~\cite{45Channel100_OL2017a}, but no 64$\times$100 GHz WDM filter has been achieved.
There are some trying to push AWGs to 64 or more channels that has resulted in dramatical insertion loss and channel isolation degradation even though the channel spacing has been compromised to 25 or 50 GHz~\cite{SiliconPhotonicArrayed_OL2022a,HighResolutionUltra_JOLT2020a,UltraCompactSilicon_IJOSTIQE2014a}.
Ring resonators have been demonstrated up to 32$\times$100 GHz WDM~\cite{LowLossLow_LPR2024a}, which already requires a ring radius $\leq 3\ \mathrm{\mu m}$ to provide sufficient free spectral range (FSR)~\cite{192ChannelMonolithically_AP2025a, LowLossLow_LPR2024a,LowLosswideFsr_2023a,SiliconBasedChip_P2023a}.
It is very challenging to extend ring-only WDM filter further to 64$\times$100 GHz by shrinking the ring to 6.4 THz FSR while maintaining a low ring roundtrip loss and sufficient ring-bus coupling required by a certain WDM passing bandwidth.

With the invention of the third-order polynomial interconnected circular (TOPIC) bend, we demonstrated the first silicon ring-based 32$\times$100 GHz WDM filter with low insertion loss and low tuning power~\cite{LowLossLow_LPR2024a}.
Here, we extend this approach to 64$\times$100 GHz WDM by cascading an MZI filter at each ring drop port, where the MZI is artfully designed such that its arm length difference is constructed using half of the ring.
This design allows precise matching of the FSR and operating wavelengths between the ring and MZI without the requirement for dynamic tuning, enabling 64$\times$100 GHz WDM over two FSRs of the ring.

Beyond simply achieving high channel counts, modern WDM systems require spectral responses with flat-top passbands that ensure uniform signal quality across all WDM channels by maintaining consistent insertion loss and minimal amplitude variation within the passband.
This is critical for optical communication systems employing high-order modulation formats, where amplitude fluctuations directly degrade signal quality and reduce overall system capacity.
In the literature, flat-top passbands in MZI filters can be achieved by cascading multiple phase-shifting stages~\cite{HighOrderLattice_2025a,CascadedMachZehnder_OE2013b,CompactLowLoss_JOLT2004a}, while flat-top responses in ring resonators can be realized using high-order ring structures~\cite{HighOrderAdiabatic_JOLT2021a,CascadingSecondOrder_JOLT2017a,VeryHighOrder_IPTL2004a}.
The proof-of-concept demonstration presented here employs single-stage phase shifters in the MZIs and single-ring configurations, resulting in peaked spectral responses.
However, the proposed ring-MZI cascade architecture is not limited to such simple configurations.
In our parallel work on building blocks for high-channel-count WDM systems, the MZI filters can be designed with multiple phase-shifting stages to achieve flat-top responses~\cite{SiliconPhotonicCwdm_2026a}, and the TOPIC ring resonators can be implemented as double-ring structures to achieve flat-top transmissions~\cite{UltraCompactSilicon_2025a}.
These approaches can be seamlessly integrated into the ring-MZI cascade architecture, paving the way for high-performance WDM systems with flat-top passbands in the future.

\begin{figure}[!b]%
    \centering
    \includegraphics{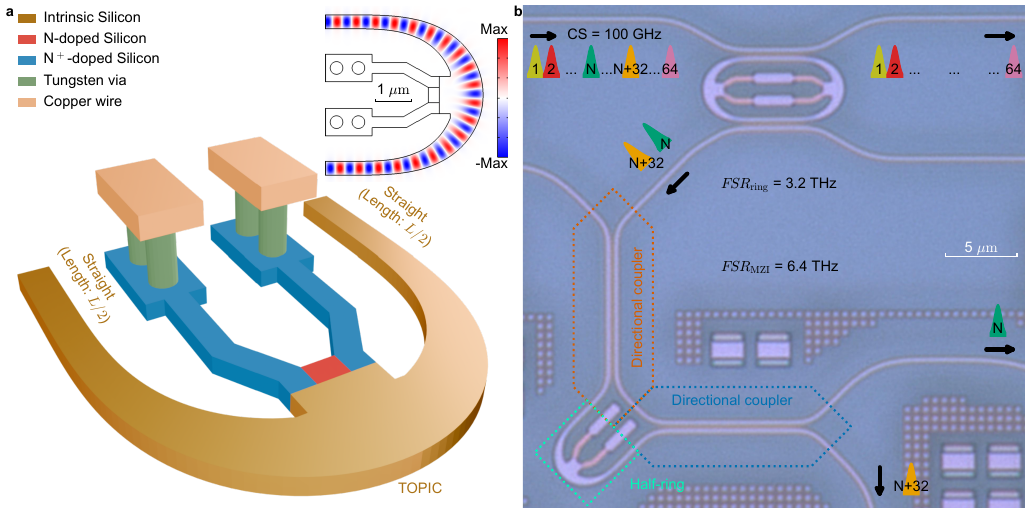}
    \caption{(a) The schematic of the half-ring waveguide implemented with one 180$^\circ$ TOPIC bend and two identical straight segments.
    The inset is the top view of the half-ring waveguide with the simulated optical field distribution, magnetic field along thickness direction at the waveguide center plane.
    In this simulation, the material stacks are SOI with silicon oxide as the top cladding, the analysis is carried out for fundamental TE mode, the Si thickness is 210 nm, the straight waveguide width is 380 nm, and the working wavelength is 1310 nm.
    (b) The microscope image of the fabricated half-ring based MZI filter to double the WDM wavelength range of the ring FSR.
    CS: channel spacing, $FSR_{\mathrm{ring}}$: the FSR of the ring, and $FSR_{\mathrm{MZI}}$: the FSR of the MZI filter.
    The TOPIC bends employed in this work feature a radius of 2 $\mathrm{\mu m}$, with transition angles of $\theta_{p,i}=43.2^{\circ}$ and $\theta_{p,o}=62.35^{\circ}$ for the inner and outer boundaries respectively.
    The fabrication is done using imec's iSiPP300 platform.}
    \label{fig_scheme}
\end{figure}

\section{Results and discussion}

The TOPIC bend exhibits zero curvature and zero curvature derivative at both endpoints, thereby minimizing mode mismatch loss when connecting to straight waveguides~\cite{LowLossLow_LPR2024a}.
This characteristic enables the construction of a low-loss half-ring waveguide comprising one 180-degree TOPIC bend and two straight segments, as illustrated in Fig.~\ref{fig_scheme}a.
As demonstrated in the inset, light propagates through the TOPIC wide-width region as a whispering gallery mode, which is exploited to embed a doped silicon heater for high thermal tuning efficiency without introducing additional loss.
Using the half-ring waveguide as the fundamental building block, a ring resonator can be readily constructed from two identical units.
The straight segment length $L$ in the half-ring waveguide can be adjusted to tune the ring resonance wavelengths and FSR.
To develop a WDM filter operating over two FSRs of the ring, an MZI filter can be cascaded at the ring drop port.
The critical challenge of this architecture lies in precisely aligning the MZI interference peaks with the ring resonance wavelengths while ensuring the MZI FSR is exactly double that of the ring.
To address this challenge, we propose an elegant MZI design as illustrated in Fig.~\ref{fig_scheme}b.
The MZI filter consists of one half-ring waveguide (indicated by the cyan rectangle) and two identical directional couplers (marked by the orange and blue hexagons).
Notably, the directional couplers feature a specialized design for the two ports connected to the MZI arms, which are positioned at a 90-degree angle to each other.
When connected as in Fig.~\ref{fig_scheme}b, the path length difference between the MZI arms is solely determined by the half-ring waveguide.
Consequently, the MZI FSR is precisely double of the ring FSR, with interference wavelengths theoretically aligned with the ring resonance wavelengths.
\begin{figure}[!b]
    \centering
    \includegraphics{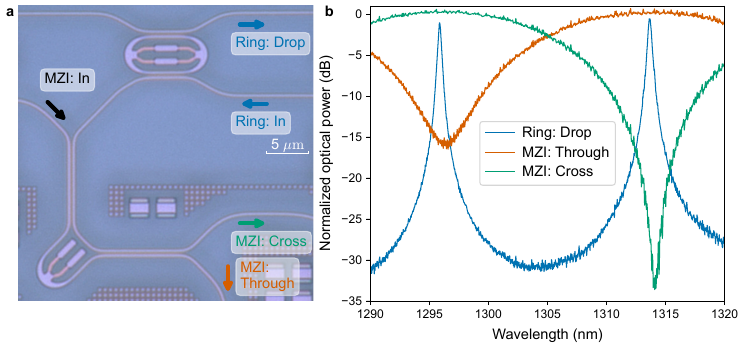}
    \caption{The measurement port configurations (a) and the measured transmission spectra (b) of the ring and MZI.}
    \label{fig_ring_mz_comparison}
\end{figure}
However, in practical fabrication, a slight operating wavelength misalignment occurs when the ring and MZI employ identical half-ring waveguide designs, as reported in our recent conference proceedings~\cite{SiliconRingBased_2024a}.
This misalignment is primarily attributed to the coupling between the ring and bus waveguide, which induces a minor perturbation in the effective refractive index of the ring waveguide, as well as in the lithography and etching process conditions of the respective local regions.
To address this issue, we slightly reduced the straight segment length of the half-ring waveguide in the MZI $L_{MZI}$ compared to the ring $L_{ring}$ with $L_{ring} - L_{MZI} = 50\ \mathrm{nm}$.
To characterize the individual transmission spectra of the ring and MZI filter, the two unused ports by the WDM filter, marked as MZI:In and Ring:In in Fig.~\ref{fig_ring_mz_comparison}a, are used as the inputs for the MZI and ring measurements respectively.
As shown in Fig.~\ref{fig_ring_mz_comparison}b, the measured spectra indicate that the MZI interference peaks are well aligned with the ring resonance wavelengths and the MZI FSR is double that of the ring without the requirement for dynamic tuning.

\begin{figure}[!b]
    \centering
    \includegraphics{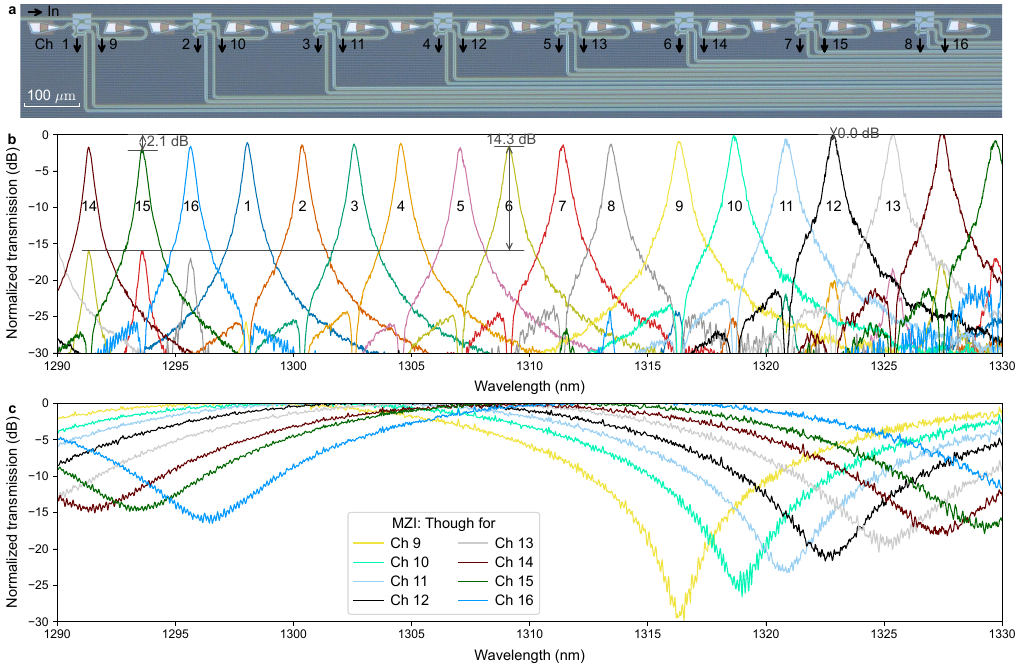}
    \caption{ The microscope image (a), the measured WDM transmission spectra (b), and the corresponding MZI Through spectra (c) of the proposed WDM 16$\times$400 GHz filter.}
    \label{fig_wdm16x400}
\end{figure}

\begin{figure}[!b]
    \centering
    \includegraphics{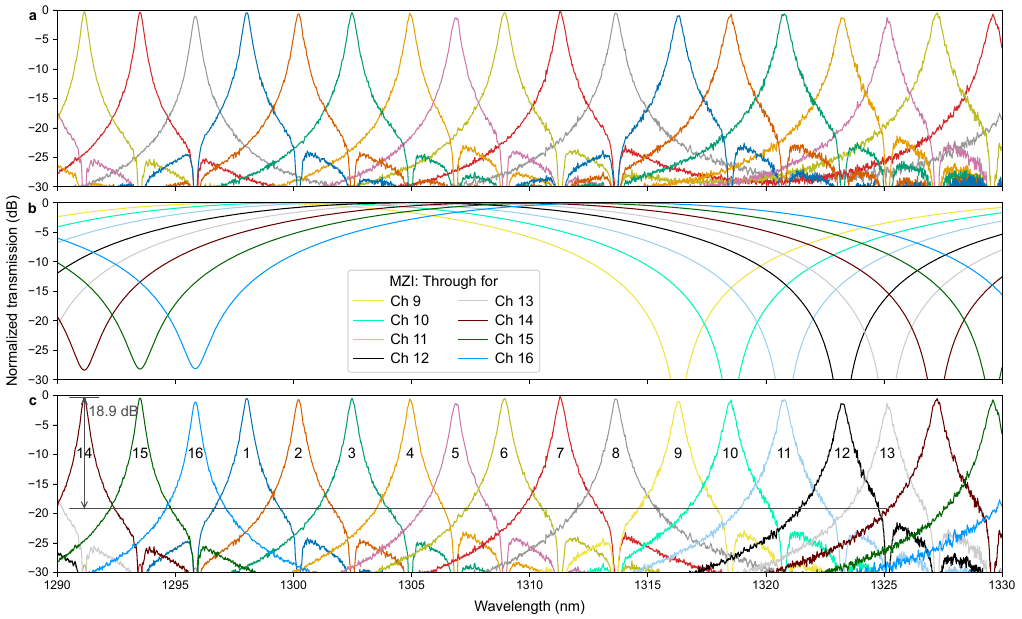}
    \caption{ The measured transmission spectra of the individual rings with the port configuration shown in Fig.~\ref{fig_ring_mz_comparison}a (a), the synthesized MZI through spectra with bent directional couplers~\cite{LowLossSilicon_JOLT2024a} (b), and the synthesized WDM 16$\times$400 GHz spectra by combining these ring and MZI transmissions (c).
    }
    \label{fig_wdm16x400_synthesized}
\end{figure}

\begin{figure}[!b]
    \centering
    \includegraphics{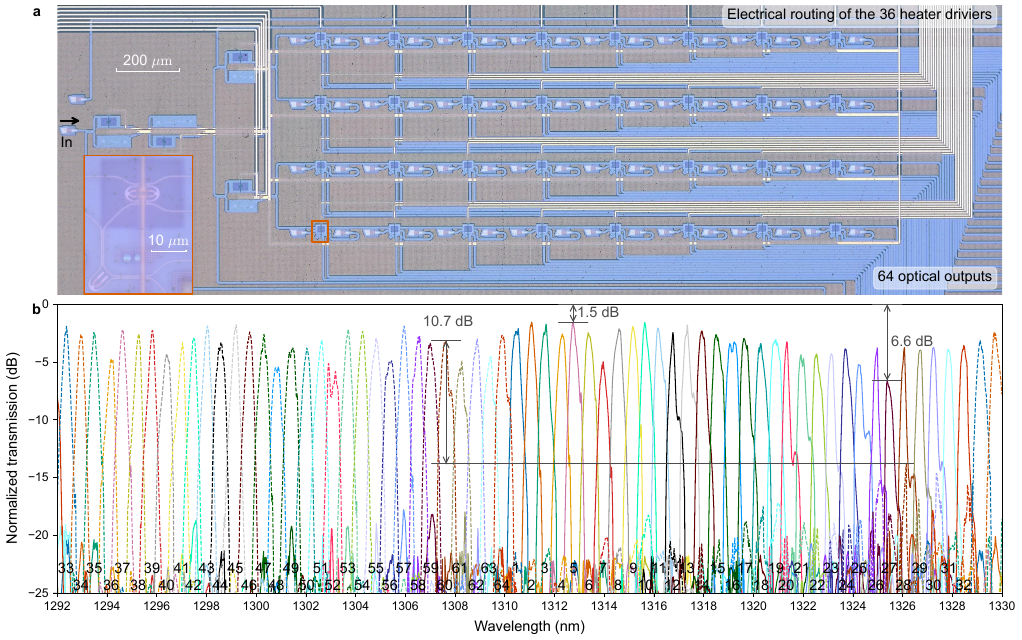}
    \caption{ The microscope image (a) and the measured transmission spectra (b) of the proposed WDM 64$\times$100 GHz filter.
    The inset is the zoomed-in view of one ring-MZI filter unit.}
    \label{fig_wdm64x100}
\end{figure}

\begin{figure}[!b]
    \centering
    \includegraphics{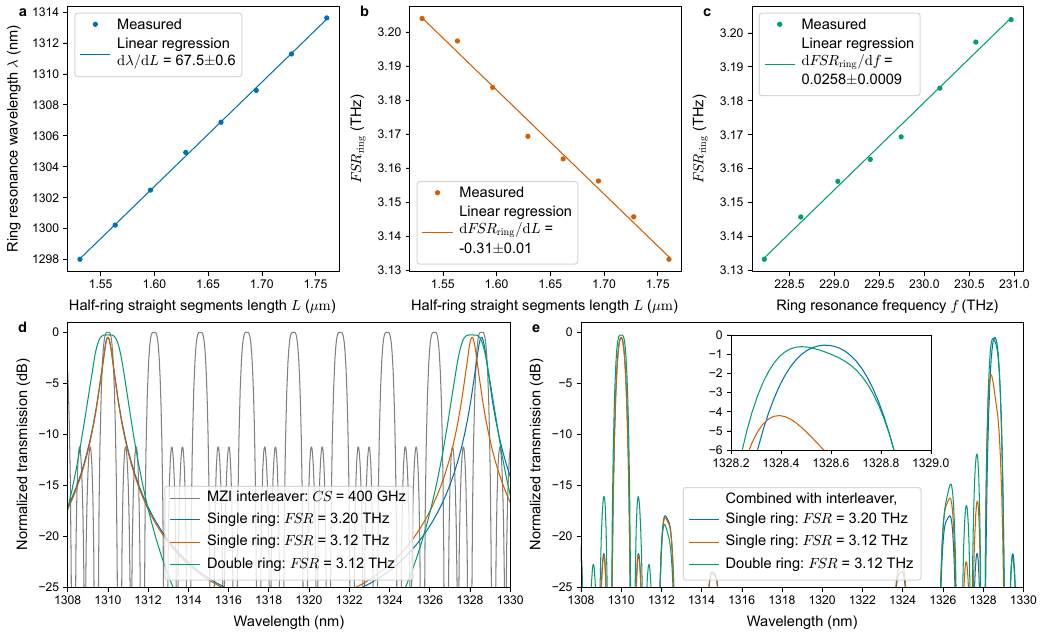}
    \caption{ The measured ring resonance wavelength (a) and the FSR (b) under different half-ring straight segments length, and the linear regression analysis between the FSR and the resonance frequency (c).
    The calculated individual (d) and the combined (e) MZI interleaver and ring dropping spectra based on their analytical models.}
    \label{fig_fsr_dispersion}
\end{figure}

Eight ring-MZI blocks, with the half-ring straight segment lengths set for 400 GHz channel spacing, are cascaded to form a 16$\times$400 GHz WDM filter, as shown in Fig.~\ref{fig_wdm16x400}a.
The measured transmission spectra are presented in Fig.~\ref{fig_wdm16x400}b.
Without requiring dynamic tuning, all 16 channels are well-defined with a channel spacing of 396$\pm$3 GHz, a full width at half maximum (FWHM) of 68$\pm$8 GHz, an insertion loss of 1.3$\pm$0.6 dB (ranging from 0.0 to 2.1 dB), and channel isolation $\geq$14.3 dB.
The channel isolation is primarily limited by the wavelength-dependent performance of the directional couplers employed in the MZIs.
These straight waveguide-based directional couplers are designed for a 0.5:0.5 splitting ratio at 1310 nm.
However, the splitting ratio deviates from the ideal 0.5:0.5 value with an approximately linear dependence on wavelength shift, resulting in degraded extinction ratio in the MZI through transmission~\cite{EnhancedOperationRange_IPTL2025a}.
As verified by Fig.~\ref{fig_wdm16x400}c, the extinction ratio of the MZI through transmission decreases progressively as the wavelength deviates from 1310 nm.
To improve channel isolation in future designs, directional couplers incorporating TOPIC bends can be employed, which provide significantly less wavelength-dependent coupling variation than straight waveguide-based couplers~\cite{LowLossSilicon_JOLT2024a}.
This bent directional coupler design can significantly enhance the MZI extinction ratio, improving the worst-case value from 14.3 dB (Fig.~\ref{fig_wdm16x400}c) to 28.1 dB (Fig.~\ref{fig_wdm16x400_synthesized}b).
When these bent directional couplers are incorporated into the MZI, the channel isolation can be improved to $\geq$18.9 dB as indicated by the synthesized WDM 16$\times$400 GHz spectra in Fig.~\ref{fig_wdm16x400_synthesized}c.
To further improve the channel isolation, cascaded ring resonators can be employed~\cite{UltraCompactSilicon_2025a}.

To achieve 64$\times$100 GHz WDM, a 4-channel interleaver~\cite{LowLossLow_LPR2024a} is employed to demultiplex the 64$\times$100 GHz input into four groups of 16$\times$400 GHz signals.
Each group is subsequently processed by the demonstrated ring-MZI WDM 16$\times$400 GHz filter, as illustrated in Fig.~\ref{fig_wdm64x100}a.
The four ring-MZI filters are designed with 100 GHz staggered operating wavelengths by adjusting the half-ring straight segment lengths to accommodate a 64$\times$100 GHz WDM grid.
To dynamically align the WDM grid between the interleaver and the four ring-MZI filters, doped silicon heaters are integrated into both the interleaver and rings~\cite{LowLossLow_LPR2024a}.
No heater is integrated into the half-ring based MZIs as they exhibit sufficiently wide passbands, benefiting from the large FSR of 6.4 THz.
As shown in Fig.~\ref{fig_ring_mz_comparison}a, the MZI exhibits extinction ratios $\geq$ 15.7 (24.9) dB within $\pm$100 GHz, corresponding to the spectral range of 2 channels, around the through (cross) transmission peak.
Therefore, dynamic tuning for the half-ring based MZIs is not required for WDM grid alignment.
The measured spectra after alignment are shown in Fig.~\ref{fig_wdm64x100}b.
This WDM filter demonstrates 64 channels with a channel spacing of 100$\pm$1 GHz, channel FWHM of 50$\pm$10 GHz, insertion loss of 3.2$\pm$1.1 dB (ranging from 1.5 to 6.6 dB), and channel isolation $\geq$10.7 dB.
Compared to silicon WDM filters reported in the literature with 64 or more channels (Tab.~\ref{tab_wdm_comparison}), the results presented here show significant improvement in terms of insertion loss and channel isolation.
The channel isolation here is primarily limited by the same issue of wavelength-dependent directional coupling as in the 16$\times$400 GHz WDM filter, which can be further improved by employing bent directional couplers as discussed above.
Moreover, the potential of the proposed ring-MZI WDM architecture has not been fully realized yet.

First, the footprint of this proof-of-concept 64$\times$100 GHz WDM filter is $\sim$ 2.5$\times$0.7 $\mathrm{mm^2}$, which is relatively large.
As shown in Fig.~\ref{fig_wdm16x400}a and \ref{fig_wdm64x100}a, most of the footprint is occupied by the additional input and output grating couplers for individual ring and MZI characterization purposes, which are not necessary in practical applications.
The essential footprint of every 2 channels in this architecture is only $\sim$0.03$\times$0.04 $\mathrm{mm^2}$ as shown in the inset of Fig.~\ref{fig_wdm64x100}a.
Upon removing these additional grating couplers and bringing each row of rings closer together for collective tuning~\cite{SiliconRingBased_OFCC22026a}, the footprint of the 64$\times$100 GHz WDM filter can be reduced to approximately 0.8$\times$0.7 $\mathrm{mm^2}$.

Second, the insertion loss of the 64$\times$100 GHz WDM filter has degraded significantly compared to the 16$\times$400 GHz WDM filter, even though they are implemented using the same technology.
The main reason for this degradation is the FSR misalignment between the interleaver and the rings originated from the dispersion of the ring FSR.
To regulate the ring resonance wavelengths to cover the entire 64$\times$100 GHz grid, the half-ring straight segment length $L$ is adjusted.
As shown in Fig.~\ref{fig_fsr_dispersion}a, the ring resonance wavelengths redshift linearly with increasing $L$.
Meanwhile, the FSR decreases linearly with $L$ as depicted in Fig.~\ref{fig_fsr_dispersion}b.
Therefore, the ring FSR decreases with increasing resonance frequency, as illustrated in Fig.~\ref{fig_fsr_dispersion}c.
The maximum resonance frequency shift required here is 3.1 THz, which results in an FSR variation of $\sim$0.08 THz.
Fig.~\ref{fig_fsr_dispersion}d shows the calculated transmission spectra of the ring and MZI based on their analytical models, and Fig.~\ref{fig_fsr_dispersion}e presents the combined spectra.
As shown in the zoomed-in inset, the 0.08 THz FSR mismatch causes the insertion loss to increase from 0.54 dB to 4.21 dB.
The measured channel insertion loss ranges from 1.5 dB to 6.6 dB, primarily because the rings employed in the 64$\times$100 GHz WDM filter exhibit FSRs spanning from 3.20 THz to 3.12 THz.
This issue can be mitigated by replacing the single ring with a double-ring structure~\cite{UltraCompactSilicon_2025a}, which maintains a low insertion loss of 0.62 dB even in the presence of a 0.08 THz FSR mismatch.
It is worth noting that the maximum FSR mismatch can be halved if the median-sized ring is configured to a 3.2 THz FSR, rather than aligning the smallest ring to 3.2 THz FSR as implemented in this demonstration.
To further extend this architecture to wider WDM spectral ranges, such as spanning 4 ring FSRs, a general recommendation would be to configure the ring with a flat-top dropping response that covers the required passband plus the maximum FSR mismatch.
In this case, higher-order ring structures should be employed, while single-ring configurations should be avoided due to their Lorentzian-shaped dropping response.

Third, the passband of each channel can be further engineered to achieve a flat-top spectral shape.
The proposed architecture forms each passband through four cascaded filters: two dense wavelength division multiplexing (DWDM) MZI filters in the interleaver, one ring resonator, and one coarse wavelength division multiplexing (CWDM) MZI filter.
In this proof-of-concept demonstration, the second DWDM MZI filter and the CWDM MZI filter are implemented with single-stage phase shifters that produce sinusoidal transmission characteristics, while the ring is implemented as a single add-drop configuration with Lorentzian-shaped filtering response.
The observed WDM channel passband results from the multiplexing of these four filters, yielding a peaked spectral shape.
Moreover, side lobes are observed in the channel passband, as shown in Fig.~\ref{fig_wdm64x100}b, which originate from slight FSR mismatches and central wavelength misalignments among these four filters.
It is important to note that the proposed architecture is not limited to simple MZI and single-ring configurations.
All four cascaded filters in this architecture can be designed to exhibit flat-top transmission spectra, thereby enabling WDM channels with flat-top passbands and minimized side lobes.
For instance, the two DWDM MZI filters in the interleaver can be designed with two or three phase-shifting stages~\cite{CascadedMachZehnder_OE2013a}, the ring can be implemented as a double-ring structure~\cite{UltraCompactSilicon_2025a}, and the CWDM MZI filter can be extended to a lattice filter configuration~\cite{SiliconPhotonicCwdm_2026a}.

Finally, heaters can be integrated into the half-ring-based MZIs depending on the application requirements.
In this demonstration, heaters are integrated only into the interleaver and rings, which is sufficient to align each filter channel to the nearest WDM grid.
This configuration is suitable for applications where all channels collectively transmit data between two physical locations, typically referred to as Wide-and-Slow (WaS) protocols~\cite{HeterogeneousIntegrationCo_IJOEASTICAS2025a,MosaicBreakingOptics_2025a}.
In WaS applications, assigning each filter channel to a specific WDM grid is unnecessary because the data sequence information can be encoded in the data stream.
Without locking each channel to a specific wavelength, the transmitted data from all channels can still be correctly recombined, resulting in approximately half the tuning power consumption.
However, this configuration is insufficient for applications where each WDM channel must be routed to a dedicated physical destination, such as in long-haul optical communications.
In such scenarios, the half-ring-based MZIs can be designed with flat-top passbands sufficiently wide to cover the required tuning range, or be equipped with heaters, thereby enabling each WDM filter channel to be tuned to any designated wavelength.

\begin{table}[!t]
    \centering
    \caption{Performance comparison of silicon WDM filters with 64 or more channels.}
    \begin{tabular}{cccccc}
    \hline
        Reference & Channel amount & Channel spacing & Insertion loss & Channel isolation & Footprint \\
        \ & \ & (GHz) & (dB) & (dB) & ($\mathrm{mm^2}$)\\ \hline
        \cite{SiliconPhotonicArrayed_OL2022a} & 64 & 50 & $\sim$5 &$\geq 10$ & 2.3$\times$2 \\ 
        \cite{HighResolutionUltra_JOLT2020a} & 64 & 50 & 5.6$\sim$8 & $\geq 9$ & 0.65$\times$1.06 \\ 
        \cite{UltraCompactSilicon_IJOSTIQE2014a} & 512 & 25 & $\sim$45 & $\sim$4 & 11$\times$16 \\  
        This work & 64 & 100 & 3.2$\pm$1.1 & $\geq 10.7$ & 2.5$\times$0.7 \\
        \ & \ & \ & \ & \ & ($\sim$0.8$\times$0.7)*\\\hline
    \end{tabular}
    \par\noindent *Excluding redundant grating couplers used for individual component characterization.
    \label{tab_wdm_comparison}
\end{table}
\section{Conclusion}

In this work, we have demonstrated, to the best of our knowledge, the first silicon 64$\times$100 GHz WDM filter using a novel ring-MZI cascade architecture.
Building upon our TOPIC ring-based 32$\times$100 GHz WDM filter~\cite{LowLossLow_LPR2024a}, the key advancement presented here is the integration of MZIs at the ring drop ports to double the WDM spectral range.
We have not found any prior art employing this ring-MZI cascade architecture for WDM filters.
However, a similar cascading with a different FSR relationship has been proposed to enhance sensitivity in sensing applications~\cite{UltraHighSensitivity_SAABC2013a}.
Compared to the existing ring-MZI cascading, the primary innovation lies in the MZI layout, where the arm length difference is constructed entirely from half of the ring structure.
This geometric constraint ensures that the MZI FSR is exactly double that of the ring resonator with aligned operating wavelengths, enabling WDM operation over two FSRs of the ring without requiring dynamic tuning alignment between the ring and the MZI, as experimentally verified with a 16$\times$400 GHz WDM filter.
Building upon this design, we demonstrated a 64$\times$100 GHz WDM filter achieving a channel spacing of 100$\pm$1 GHz, insertion loss of 3.2$\pm$1.1 dB, and channel isolation $\geq$10.7 dB, representing significant improvements over existing silicon WDM filters with 64 or more channels.
Moreover, the proposed ring-MZI WDM architecture exhibits excellent potential for achieving substantially lower insertion loss, higher channel isolation, and reduced footprint.
We identified ring FSR dispersion as the primary cause of insertion loss degradation when scaling from 16$\times$400 GHz to 64$\times$100 GHz operation, which can be mitigated using double-ring structures in future designs.
Additionally, the wavelength-dependent coupling variation of the directional couplers in the MZIs limits channel isolation, which can be improved by incorporating broadband directional couplers.
The footprint of this proof-of-concept 64$\times$100 GHz WDM filter can be significantly reduced by eliminating the additional grating couplers used for individual component characterization, resulting in an essential footprint of only $\sim$0.03$\times$0.04 $\mathrm{mm^2}$ per two channels.
All in all, this work establishes a promising pathway for scaling silicon WDM systems beyond conventional limits, enabling higher-capacity optical communication networks required for emerging artificial intelligence and machine learning applications.

\section{Methods}

\textbf{Device fabrication.} The devices were fabricated using imec's iSiPP300 silicon photonics platform on 300 mm wafers with 220 nm thick silicon-on-insulator (SOI).
The UCUT process, which removes the silicon substrate beneath the SOI, was employed to form the doped silicon heater using the same design as in our previous work~\cite{LowLossLow_LPR2024a}, achieving tuning powers of 1.50 $\mathrm{mW/\pi}$ for the heaters in the interleaver and 5.85 $\mathrm{mW/\pi}$ for the rings.

\textbf{Spectra synthesis in Fig.~\ref{fig_wdm16x400_synthesized}.} The individual ring transmission spectra were obtained from measurements using the port configuration shown in Fig.~\ref{fig_ring_mz_comparison}a. 
The MZI spectra were synthesized based on the directional coupling model described in Ref.~\cite{LowLossSilicon_JOLT2024a}, with the arm length differences set to match the corresponding ring resonances.
The final WDM spectra were generated by multiplying the measured ring drop transmission with the synthesized MZI through or cross transmission.

\textbf{Analytical model of the interleaver for Fig.~\ref{fig_fsr_dispersion}.} The 4-channel interleaver is constructed by cascading two lattice filters with the same design as described in our previous work~\cite{LowLossLow_LPR2024a}.
The first lattice filter in the cascade comprises two phase-shifting stages to achieve flat-top transmission spectra~\cite{CascadedMachZehnder_OE2013a}.
The second phase-shifting stage features an arm relative length that is double that of the first stage.
The two stages are interconnected by three directional couplers with power cross-coupling ratios of 0.5, 0.29, and 0.08, respectively.
The second lattice filter consists of one phase-shifting stage connected by two directional couplers with power cross-coupling ratios of 0.5.
The arm length differences of the first and second lattice filters are configured to achieve FSRs of 200 and 400 GHz, respectively, while ensuring that 1310 nm corresponds to one of the operating wavelengths.
The losses and wavelength dependence of the directional couplers are neglected in the analytical calculations.

\textbf{Analytical model of the rings for Fig.~\ref{fig_fsr_dispersion}.} The analytical model of the rings is based on the transfer matrix method~\cite{UniversalRelationsCoupling_EL2000a}.
The ring roundtrip losses are set to 0.05 dB and the power cross-coupling ratio between the ring and bus waveguide is set to 0.09 for the single ring according to the measured data in Ref.~\cite{LowLossLow_LPR2024a}.
The two rings in the double-ring structure are assumed to be identical with the same roundtrip loss of 0.05 dB.
The rings are coupled to the add/drop bus waveguides with a power cross-coupling ratio of 0.3.
The power cross-coupling ratio between the two rings is set to 0.033 to achieve flat-top transmission spectra~\cite{CascadingSecondOrder_JOLT2017a}.
The ring roundtrip length is adjusted to achieve FSRs of 3.20 or 3.12 THz while ensuring that 1310 nm corresponds to one of the resonance wavelengths.
The losses and wavelength dependence of the ring-bus and ring-ring coupling are neglected in the analytical calculations.
\begin{acknowledgement}
This work was supported by imec's industry-affiliation R\&D program
“Optical I/O”.
\end{acknowledgement}
\bibliography{topic_ring_wdm64_ref}

@InProceedings{LowLosswideFsr_2023a,
  author     = {Novick, Asher and Jang, Kaylx and Rizzo, Anthony and Parsons, Robert and Bergman, Keren},
  booktitle  = {OFC},
  title      = {Low-LossWide-FSR Miniaturized RacetrackStyle Microring Filters for $\geq$ 1Tbps DWDM},
  year       = {2023},
  pages      = {Th3A.3},
  publisher  = {Optica Publishing Group},
  series     = {OFC},
  volume     = {29},
  collection = {OFC},
  doi        = {10.1364/ofc.2023.th3a.3},
  journal    = {Optical Fiber Communication Conference (OFC) 2023},
}

@Article{HighOrderAdiabatic_JOLT2021a,
  author   = {Liu, Dajian and Zhang, Long and Tan, Ying and Dai, Daoxin},
  journal  = {Journal of Lightwave Technology},
  title    = {High-{Order} {Adiabatic} {Elliptical}-{Microring} {Filter} with an {Ultra}-{Large} {Free}-{Spectral}-{Range}},
  year     = {2021},
  issn     = {1558-2213},
  month    = sep,
  number   = {18},
  pages    = {5910--5916},
  volume   = {39},
  doi      = {10.1109/JLT.2021.3091724},
}

@Article{SiliconBasedChip_P2023a,
  author    = {Yin, Yuxiang and Yu, Hang and Tu, Donghe and Huang, Xingrui and Yu, Zhiguo and Guan, Huan and Li, Zhiyong},
  journal   = {Photonics},
  title     = {A {Silicon}-{Based} {On}-{Chip} 64-{Channel} {Hybrid} {Wavelength}- and {Mode}-{Division} (de){Multiplexer}},
  year      = {2023},
  issn      = {2304-6732},
  month     = feb,
  number    = {2},
  pages     = {183},
  volume    = {10},
  copyright = {http://creativecommons.org/licenses/by/3.0/},
  doi       = {10.3390/photonics10020183},
  language  = {en},
  publisher = {Multidisciplinary Digital Publishing Institute},
  url       = {https://www.mdpi.com/2304-6732/10/2/183},
  urldate   = {2023-02-14},
}

@Article{UniversalRelationsCoupling_EL2000a,
  author   = {Yariv, A.},
  journal  = {Electronics Letters},
  title    = {Universal relations for coupling of optical power between microresonators and dielectric waveguides},
  year     = {2000},
  issn     = {1350-911X},
  month    = feb,
  number   = {4},
  pages    = {321--322},
  volume   = {36},
  doi      = {10/dkfkfx},
  language = {en},
  urldate  = {2021-06-22},
}

@InProceedings{SiliconRingBased_OFCC22026a,
  author       = {Deng, Qingzhong and Kobbi, Hakim and De Coster, Jeroen and Magdziak, Rafal and Jakanadan, Shalini and Balakrishnan, Sadhishkumar and Singh, Neha and Filipcic, Marko Ersek and Chakrabarti, Maumita and Velenis, Dimitrios and De Heyn, Vincent and Verheyen, Peter and Sar, Huseyin and Absil, Philippe and Ferraro, Filippo and Ban, Yoojin and Van Campenhout, Joris},
  booktitle    = {Optical Fiber Communication Conference ({OFC})},
  title        = {Silicon Ring-Based WDM Filter with a Low Tuning Power of 3.80 mW/pi per Channel},
  year         = {2026},
  pages        = {Tu3C},
  publisher    = {Optica Publishing Group},
  series       = {OFC},
  collection   = {OFC},
  journal      = {Optical Fiber Communication Conference 2026},
  primaryclass = {physics.optics},
}

@Article{LowLossLow_LPR2024a,
  author    = {Deng, Qingzhong and El-Saeed, Ahmed H. and Elshazly, Alaa and Lepage, Guy and Marchese, Chiara and Neutens, Pieter and Kobbi, Hakim and Magdziak, Rafal and De Coster, Jeroen and Vaskasi, Javad Rahimi and Kim, Minkyu and Tong, Yeyu and Singh, Neha and Filipcic, Marko Ersek and Van Dorpe, Pol and Croes, Kristof and Chakrabarti, Maumita and Velenis, Dimitrios and De Heyn, Peter and Verheyen, Peter and Absil, Philippe and Ferraro, Filippo and Ban, Yoojin and Van Campenhout, Joris},
  journal   = {Laser \& Photonics Reviews},
  title     = {Low-Loss and Low-Power Silicon Ring Based WDM 32$\times$100 GHz Filter Enabled by a Novel Bend Design},
  year      = {2024},
  issn      = {1863-8880},
  month     = nov,
  number    = {5},
  volume    = {19},
  doi       = {10.1002/lpor.202401357},
  publisher = {Wiley},
}

@Article{SiliconPhotonicArrayed_OL2022a,
  author    = {Liu, Yingjie and Wang, Xi and Yao, Yong and Du, Jiangbing and Song, Qinghai and Xu, Ke},
  journal   = {Optics Letters},
  title     = {Silicon photonic arrayed waveguide grating with 64 channels for the 2 $\mu$m spectral range},
  year      = {2022},
  issn      = {1539-4794},
  month     = feb,
  number    = {5},
  pages     = {1186},
  volume    = {47},
  doi       = {10.1364/ol.452476},
  publisher = {Optica Publishing Group},
}

@Article{CrosstalkFree32_JOLT2023a,
  author    = {Akiyama, Tomoyuki and Nishizawa, Motoyuki and Sugama, Akio and Nakasha, Yasuhiro and Tanaka, Shinsuke and Tanaka, Yu and Hoshida, Takeshi},
  journal   = {Journal of Lightwave Technology},
  title     = {Crosstalk-Free 32-ch DWDM Demultiplexer On Standard Si Pic Platform Enabled By Fully-Integrated Cascaded AMZ Triplet},
  year      = {2023},
  issn      = {1558-2213},
  month     = feb,
  number    = {3},
  pages     = {848--854},
  volume    = {41},
  doi       = {10.1109/jlt.2022.3216024},
  publisher = {Institute of Electrical and Electronics Engineers (IEEE)},
}

@Article{UltraCompactSilicon_IJOSTIQE2014a,
  author    = {Cheung, Stanley and Tiehui Su and Okamoto, Katsunari and Yoo, S. J. B.},
  journal   = {IEEE Journal of Selected Topics in Quantum Electronics},
  title     = {Ultra-Compact Silicon Photonic 512 × 512 25 GHz Arrayed Waveguide Grating Router},
  year      = {2014},
  issn      = {1558-4542},
  month     = jul,
  number    = {4},
  pages     = {310--316},
  volume    = {20},
  doi       = {10.1109/jstqe.2013.2295879},
  publisher = {Institute of Electrical and Electronics Engineers (IEEE)},
}

@Article{HighResolutionUltra_JOLT2020a,
  author    = {Zou, Jun and Ma, Xiao and Xia, Xiang and Hu, Jinhua and Wang, Changhui and Zhang, Ming and Lang, Tingting and He, Jian-Jun},
  journal   = {Journal of Lightwave Technology},
  title     = {High Resolution and Ultra-Compact On-Chip Spectrometer Using Bidirectional Edge-Input Arrayed Waveguide Grating},
  year      = {2020},
  issn      = {1558-2213},
  month     = aug,
  number    = {16},
  pages     = {4447--4453},
  volume    = {38},
  doi       = {10.1109/jlt.2020.2992905},
  publisher = {Institute of Electrical and Electronics Engineers (IEEE)},
}

@InProceedings{LowLossHigh_2012a,
  author     = {Cheung, S. T. S. and Guan, B. and Djordjevic, S. S. and Okamoto, K. and Yoo, S. J. B.},
  booktitle  = {Conference on Lasers and Electro-Optics 2012},
  title      = {Low-loss and High Contrast Silicon-on-Insulator (SOI) Arrayed Waveguide Grating},
  year       = {2012},
  pages      = {CM4A.5},
  publisher  = {OSA},
  series     = {CLEO_SI},
  collection = {CLEO_SI},
  doi        = {10.1364/cleo_si.2012.cm4a.5},
}

@Article{CompactSingleChip_OE2011a,
  author    = {Feng, Dazeng and Feng, Ning-Ning and Kung, Cheng-Chih and Liang, Hong and Qian, Wei and Fong, Joan and Luff, B. Jonathan and Asghari, Mehdi},
  journal   = {Optics Express},
  title     = {Compact single-chip VMUX/DEMUX on the silicon-on-insulator platform},
  year      = {2011},
  issn      = {1094-4087},
  month     = mar,
  number    = {7},
  pages     = {6125},
  volume    = {19},
  doi       = {10.1364/oe.19.006125},
  publisher = {Optica Publishing Group},
}

@Article{45Channel100_OL2017a,
  author    = {Li, Kai-li and Zhang, Jia-shun and An, Jun-ming and Li, Jian-guang and Wang, Liang-liang and Wang, Yue and Wu, Yuan-da and Yin, Xiao-jie and Hu, Xiong-wei},
  journal   = {Optoelectronics Letters},
  title     = {A 45-channel 100 GHz AWG based on Si nanowire waveguides},
  year      = {2017},
  issn      = {1993-5013},
  month     = may,
  number    = {3},
  pages     = {161--164},
  volume    = {13},
  doi       = {10.1007/s11801-017-7051-4},
  publisher = {Springer Science and Business Media LLC},
}

@Article{192ChannelMonolithically_AP2025a,
  author    = {Peng, Yingying and Zhao, Weike and Yi, Xiaolin and Liu, Dajian and Xiang, Yuluan and Wan, Yuanjian and Li, Kang and Wang, Jian and Shi, Yaocheng and Dai, Daoxin},
  journal   = {ACS Photonics},
  title     = {192-Channel Monolithically Integrated Reconfigurable Optical Add-Drop Multiplexer on Silicon for Ultrahigh-Capacity Hybrid WDM-PDM-MDM Systems},
  year      = {2025},
  issn      = {2330-4022},
  month     = aug,
  doi       = {10.1021/acsphotonics.5c00429},
  publisher = {American Chemical Society (ACS)},
}

@InProceedings{SiliconRingBased_2024a,
  author    = {Deng, Qingzhong and De Coster, Jeroen and Magdziak, Rafal and El-Saeed, Ahmed H. and Elshazly, Alaa and Lepage, Guy and Marchese, Chiara and Kobbi, Hakim and Singh, Neha and Filipcic, Marko Ersek and Croes, Kristof and Velenis, Dimitrios and Chakrabarti, Maumita and De Heyn, Peter and Verheyen, Peter and Absil, Philippe and Ferraro, Filippo and Ban, Yoojin and Van Campenhout, Joris},
  booktitle = {2024 IEEE Silicon Photonics Conference (SiPhotonics)},
  title     = {Silicon Ring Based Wavelength Division Multiplexing With an Ultra-wide Spectral Range of 6.4 THz},
  year      = {2024},
  month     = apr,
  pages     = {1--2},
  publisher = {IEEE},
  doi       = {10.1109/siphotonics60897.2024.10543750},
}

@Article{EnhancedOperationRange_IPTL2025a,
  author    = {Bayoumi, Ahmed and Oktay, Mehmet and Elshazly, Alaa and Kobbi, Hakim and Magdziak, Rafal and Lepage, Guy and Marchese, Chiara and Rahimi Vaskasi, Javad and Bipul, Swetanshu and Bode, Dieter and Velenis, Dimitrios and Chakrabarti, Maumita and Verheyen, Peter and Absil, Philippe and Ferraro, Filippo and Ban, Yoojin and Van Campenhout, Joris and Bogaerts, Wim and Deng, Qingzhong},
  journal   = {IEEE Photonics Technology Letters},
  title     = {Enhanced Operation Range of Silicon MZI Filters Using a Broadband Bent Directional Coupler},
  year      = {2025},
  issn      = {1941-0174},
  month     = may,
  number    = {9},
  pages     = {500--503},
  volume    = {37},
  doi       = {10.1109/lpt.2025.3553059},
  publisher = {Institute of Electrical and Electronics Engineers (IEEE)},
}

@Article{LowLossSilicon_JOLT2024a,
  author    = {El-Saeed, Ahmed H. and Elshazly, Alaa and Kobbi, Hakim and Magdziak, Rafal and Lepage, Guy and Marchese, Chiara and Vaskasi, Javad Rahimi and Bipul, Swetanshu and Bode, Dieter and Filipcic, Marko Ersek and Velenis, Dimitrios and Chakrabarti, Maumita and Heyn, Peter De and Verheyen, Peter and Absil, Philippe and Ferraro, Filippo and Ban, Yoojin and Campenhout, Joris Van and Bogaerts, Wim and Deng, Qingzhong},
  journal   = {Journal of Lightwave Technology},
  title     = {Low-Loss Silicon Directional Coupler with Arbitrary Coupling Ratios for Broadband Wavelength Operation Based on Bent Waveguides},
  year      = {2024},
  issn      = {0733-8724},
  month     = sep,
  number    = {17},
  pages     = {6011-6018},
  volume    = {42},
  doi       = {10.1109/jlt.2024.3407339},
  publisher = {Institute of Electrical and Electronics Engineers (IEEE)},
}

@Article{UltraHighSensitivity_SAABC2013a,
  author    = {La Notte, Mario and Passaro, Vittorio M.N.},
  journal   = {Sensors and Actuators B: Chemical},
  title     = {Ultra high sensitivity chemical photonic sensing by Mach--Zehnder interferometer enhanced Vernier-effect},
  year      = {2013},
  issn      = {0925-4005},
  month     = jan,
  pages     = {994--1007},
  volume    = {176},
  doi       = {10.1016/j.snb.2012.10.008},
  publisher = {Elsevier BV},
}

@InProceedings{MosaicBreakingOptics_2025a,
  author     = {Benyahya, Kaoutar and Diaz, Ariel Gomez and Liu, Junyi and Lyutsarev, Vassily and Pantouvaki, Marianna and Shi, Kai and Siew, Shawn Yohanes and Ballani, Hitesh and Burridge, Thomas and Cletheroe, Daniel and Karagiannis, Thomas and Robertson, Brian and Rowstron, Ant and Yang, Mengyang and Costa, Paolo},
  booktitle  = {Proceedings of the ACM SIGCOMM 2025 Conference},
  title      = {Mosaic: Breaking the Optics versus Copper Trade-off with a Wide-and-Slow Architecture and MicroLEDs},
  year       = {2025},
  month      = aug,
  pages      = {234--247},
  publisher  = {ACM},
  series     = {SIGCOMM {\textquoteright}25},
  collection = {SIGCOMM {\textquoteright}25},
  doi        = {10.1145/3718958.3750510},
}

@Article{HeterogeneousIntegrationCo_IJOEASTICAS2025a,
  author    = {Yang, Yu-Tao and Hung, Chih-Ming},
  journal   = {IEEE Journal on Emerging and Selected Topics in Circuits and Systems},
  title     = {Heterogeneous Integration in Co-Packaged Optics},
  year      = {2025},
  issn      = {2156-3365},
  month     = sep,
  number    = {3},
  pages     = {427--437},
  volume    = {15},
  doi       = {10.1109/jetcas.2025.3590744},
  publisher = {Institute of Electrical and Electronics Engineers (IEEE)},
}

@Article{CascadedMachZehnder_OE2013a,
  author    = {Horst, Folkert and Green, William M. J. and Assefa, Solomon and Shank, Steven M. and Vlasov, Yurii A. and Offrein, Bert Jan},
  journal   = {Optics express},
  title     = {Cascaded Mach-Zehnder wavelength filters in silicon photonics for low loss and flat pass-band WDM (de-)multiplexing},
  year      = {2013},
  issn      = {1094-4087},
  month     = may,
  note      = {ZSCC: 0000280 Publisher: Optical Society of America},
  number    = {10},
  pages     = {11652--11658},
  volume    = {21},
  copyright = {\&\#169; 2013 OSA},
  doi       = {10.1364/OE.21.011652},
  language  = {EN},
  publisher = {The Optical Society},
  url       = {https://www.osapublishing.org/oe/abstract.cfm?uri=oe-21-10-11652},
  urldate   = {2021-04-10},
}

@InProceedings{HighOrderLattice_2025a,
  author     = {Teng, Min and Wu, Hao and Cheng, Ning and Zheng, Xuezhe},
  booktitle  = {Optical Fiber Communication Conference (OFC) 2025},
  title      = {A High-Order Lattice Filter with Enhanced Passband and Roll-off for CWDM4 Applications},
  year       = {2025},
  pages      = {Tu2G.3},
  publisher  = {Optica Publishing Group},
  series     = {OFC},
  collection = {OFC},
  doi        = {10.1364/ofc.2025.tu2g.3},
}

@Article{CascadedMachZehnder_OE2013b,
  author    = {Horst, Folkert and Green, William M. J. and Assefa, Solomon and Shank, Steven M. and Vlasov, Yurii A. and Offrein, Bert Jan},
  journal   = {Optics Express},
  title     = {Cascaded {Mach}-{Zehnder} wavelength filters in silicon photonics for low loss and flat pass-band {WDM} (de-)multiplexing},
  year      = {2013},
  issn      = {1094-4087},
  month     = may,
  note      = {ZSCC: 0000280 Publisher: Optical Society of America},
  number    = {10},
  pages     = {11652--11658},
  volume    = {21},
  copyright = {\&\#169; 2013 OSA},
  doi       = {10/gf33kg},
  language  = {EN},
  url       = {https://www.osapublishing.org/oe/abstract.cfm?uri=oe-21-10-11652},
  urldate   = {2021-04-10},
}

@Article{CompactLowLoss_JOLT2004a,
  author    = {Oguma, M. and Kitoh, T. and Inoue, Y. and Mizuno, T. and Shibata, T. and Kohtoku, M. and Hibino, Y.},
  journal   = {Journal of Lightwave Technology},
  title     = {Compact and Low-Loss Interleave Filter Employing Lattice-Form Structure and Silica-Based Waveguide},
  year      = {2004},
  issn      = {0733-8724},
  month     = mar,
  number    = {3},
  pages     = {895--902},
  volume    = {22},
  doi       = {10.1109/jlt.2004.824544},
  publisher = {Institute of Electrical and Electronics Engineers (IEEE)},
}

@Article{VeryHighOrder_IPTL2004a,
  author   = {Little, B.E. and Chu, S.T. and Absil, P.P. and Hryniewicz, J.V. and Johnson, F.G. and Seiferth, F. and Gill, D. and Van, V. and King, O. and Trakalo, M.},
  journal  = {IEEE Photonics Technology Letters},
  title    = {Very high-order microring resonator filters for {WDM} applications},
  year     = {2004},
  issn     = {1941-0174},
  month    = oct,
  number   = {10},
  pages    = {2263--2265},
  volume   = {16},
  doi      = {10.1109/LPT.2004.834525},
  url      = {https://ieeexplore.ieee.org/abstract/document/1336897},
  urldate  = {2024-03-07},
}

@Article{CascadingSecondOrder_JOLT2017a,
  author    = {Zhang, Lei and Zhao, Haiyang and Wang, Haoyan and Shao, Sizhu and Tian, Wenjing and Ding, Jainfeng and Fu, Xin and Yang, Lin},
  journal   = {Journal of Lightwave Technology},
  title     = {Cascading Second-Order Microring Resonators for a Box-Like Filter Response},
  year      = {2017},
  issn      = {1558-2213},
  month     = dec,
  number    = {24},
  pages     = {5347--5360},
  volume    = {35},
  booktitle = {Up and Running with DAX for Power BI},
  doi       = {10.1109/jlt.2017.2775658},
  isbn      = {9781484281888},
  publisher = {IEEE},
}

@InProceedings{UltraCompactSilicon_2025a,
  author     = {Deng, Qingzhong and De Coster, Jeroen and Magdziak, Rafal and El-Saeed, Ahmed H. and Elshazly, Alaa and Oktay, Mehmet and Marchese, Chiara and Lepage, Guy and Vaskasi, Javad Rahimi and Balakrishnan, Sadhishkumar and Singh, Neha and Bode, Dieter and Filipcic, Marko Ersek and Chakrabarti, Maumita and Velenis, Dimitrios and Verheyen, Peter and Absil, Philippe and Ferraro, Filippo and Ban, Yoojin and Van Campenhout, Joris},
  booktitle  = {Optical Fiber Communication Conference (OFC) 2025},
  title      = {Ultra-compact silicon rings with high thermal tuning efficiency demonstrated as an 8 × 400 GHz WDM filter},
  year       = {2025},
  pages      = {Tu2G.2},
  publisher  = {Optica Publishing Group},
  series     = {OFC},
  collection = {OFC},
  doi        = {10.1364/ofc.2025.tu2g.2},
}

@Article{SiliconPhotonicCwdm_2026a,
  author    = {Deng, Qingzhong and Elshazly, Alaa and Magdziak, Rafal and Witters, Liesbeth and Bipul, Swetanshu and Chakrabarti, Maumita and Velenis, Dimitrios and Ferraro, Filippo and Sar, Huseyin and Verheyen, Peter and Absil, Philippe and Ossieur, Peter and Jadli, Imene and Van Campenhout, Joris},
  journal   = {arXiv},
  title     = {Silicon Photonic CWDM Filter with Compact Footprint, Low Loss, Flat-Top Transmission and High Yield},
  year      = {2026},
  number    = {https://doi.org/10.48550/arXiv.2605.25106},
  pages     = {https://doi.org/10.48550/arXiv.2605.25106},
  copyright = {Creative Commons Attribution Non Commercial No Derivatives 4.0 International},
  doi       = {10.48550/ARXIV.2605.25106},
  publisher = {arXiv},
}

\end{document}